\documentclass[prl,floatfix,twocolumn,showpacs,amsmath,amssymb]{revtex4}
\usepackage{graphicx}
\usepackage{dcolumn}
\usepackage{bm}

\begin{document}

\title{Local moments and magnetic order in the two-dimensional Anderson-Mott 
transition}
\author{Maria Elisabetta Pezzoli,$^{1,2}$ Federico Becca,$^{1,2}$ 
Michele Fabrizio,$^{1,2,3}$ and Giuseppe Santoro$^{1,2,3}$}
\affiliation{
$^{1}$ CNR-INFM-Democritos National Simulation Centre, Trieste, Italy. \\
$^{2}$  International School for Advanced Studies (SISSA),
I-34014 Trieste, Italy \\
$^{3}$ International Centre for Theoretical Physics (ICTP), P.O. Box 586, 
I-34014 Trieste, Italy}

\date{\today}
\begin{abstract}
We study the role of electronic correlation in a disordered two-dimensional 
model by using a variational wave function that can interpolate 
between Anderson and Mott insulators. Within this approach, the Anderson-Mott
transition can be described both in the paramagnetic and in the magnetic 
sectors. In the latter case, we find evidence for the formation of local 
magnetic moments that order before the Mott transition. The charge gap opening 
in the Mott insulator is accompanied by the vanishing of the 
$\lim_{q\to 0} \overline{\langle n_q\rangle\langle n_{-q}\rangle}$ (the bar 
denoting the impurity average), which is related to the compressibility
fluctuations. The role of a frustrating (second-neighbor) hopping is also 
discussed, with a particular emphasis to the formation of metastable 
spin-glass states.
\end{abstract}

\maketitle

The combined action of electron-electron interaction and disorder is known to 
heavily influence the physical behavior of electron systems.~\cite{A&A} 
Recently, the observation of metallic behavior in high-mobility two-dimensional
electron-gas devices~\cite{kravchenko} has opened new perspectives in this 
subject, suggesting the possibility that a metallic behavior could be 
stabilized by a strong electron-electron interaction in two dimensions, 
in spite of the standard scaling theory of Anderson 
localization.~\cite{gang-of-4,lee} Such a proposal was first put forward 
theoretically by means of a weak-coupling renormalization group approach 
within a Fermi-liquid description,~\cite{finkelstein} and later developed along
similar directions.~\cite{castellani,punnose} A common feature of the above 
renormalization-group calculations is the crucial role played by the spin 
fluctuations that grow large as the renormalization group procedure is 
iterated. This tendency, which has been interpreted as signaling the emergence 
of local moments, suggests that electron-electron correlations become 
effectively very strong that, in turn, makes doubtful the validity of a 
Fermi liquid description.~\cite{stewart,miranda} 

Apart from the debated issue of a metal-insulator transition in two-dimensional
high-mobility devices,~\cite{abrahams,altshuler,pfeiffer,pfeiffer2} there are 
less controversial systems where the role of strong correlations concomitantly 
with disorder is well testified. Particularly emblematic is the case of Si:P 
and Si:B,~\cite{alexander,lohneysen} which are three-dimensional materials that 
show a {\it bona fide} metal-insulator transition. Here, the randomly 
distributed impurities form a very narrow band within the semiconducting gap. 
Since the local Coulomb repulsion is sizable compared to the width of the 
impurity band, this system is particularly suitable to investigate the 
interplay between disorder and interaction. Indeed, clear signatures of local 
magnetic moments are found in several thermodynamic 
quantities.~\cite{hirsch,paalanen1,paalanen2,lakner} Theoretically, the 
interplay of disorder and interaction is a very difficult question. 
Any approach based on a single-particle description, like unrestricted 
Hartree-Fock,~\cite{bhatt1,heidarian} can uncover the emergence of local 
moments only if spin-rotational symmetry is explicitly broken, introducing
spurious effects due to magnetism that can be dealt with using further 
approximate schemes.~\cite{bhatt,bhatt-bis} More sophisticated approaches, 
like those based on dynamical mean-field theory,~\cite{DMFT} can in principle 
manage without magnetism,~\cite{TMT,hofstetter,aguiar} 
but they usually miss important spatial correlations.

In this Letter, we will generalize the variational approach that has been
successfully used to describe the Mott transition in finite-dimensional clean
systems~\cite{capello,capello2,capello2-bis}. We will show that, for a 
half-filled disordered Hubbard model on a square lattice and when the 
variational wave function is forced to be paramagnetic, the Anderson to Mott 
insulator transition exists and it is continuous. When magnetism is allowed, 
we find two successive second order phase transitions: from a compressible 
paramagnetic Anderson insulator with local moments to a compressible magnetic 
Anderson insulator and then to an incompressible magnetic Mott insulator. 
Unlike previous unrestricted Hartree-Fock~\cite{heidarian} or Monte Carlo 
calculations,~\cite{scalettar} we do not find any evidence of an intermediate 
truly metallic behavior.  

We consider a half-filled Hubbard model on a square lattice with on-site 
disorder:
\begin{equation}\label{Ham}
{\cal H} =  \sum_{i,j,\sigma} t_{i,j}
c^{\dagger}_{i,\sigma} c^{\phantom{\dagger}}_{j,\sigma} + H.c.
+ \sum_i (\epsilon_i n_i + U\,n_{i,\uparrow} n_{i,\downarrow}),
\end{equation}
where $c^{\dagger}_{i,\sigma}$ ($c_{i,\sigma}$) creates (destroys) one 
electron at site $i$ with spin $\sigma$, 
$n_{i,\sigma}=c^{\dagger}_{i,\sigma} c^{\phantom{\dagger}}_{i,\sigma}$, and 
$n_i=\sum_\sigma n_{i,\sigma}$. $\epsilon_i$ are random on-site energies 
chosen independently at each site and uniformly distributed in $[-D,D]$. 
$t_{i,j}$ are the hopping parameters that we will consider limited either to 
nearest-, $t_{ij}=-t$, or to next-nearest-neighbor, $t_{ij}=-t^\prime$, sites. 
In the calculations we will consider 45 degree rotated clusters with 
$N=2 n^2$ sites, $n$ being an odd integer, and periodic boundary conditions, 
so that the non-interacting ground state is always non-degenerate at half 
filling.

Following the approach developed for clean systems,~\cite{capello} we define a 
variational wave function containing a Gutzwiller and a long-range Jastrow 
factor that apply to an uncorrelated state:
\begin{equation}
|\Psi \rangle = {\cal P}_G \, {\cal J} \, |\Phi_0 \rangle,
\end{equation}
where $|\Phi_0 \rangle$ is the ground state of a non-interacting
Hamiltonian with the same hopping parameters as in Eq.~(\ref{Ham}) but with 
variational spin-dependent on-site energies $\tilde{\epsilon}_{i\sigma}$ 
to be determined by minimizing the total energy. A paramagnetic wave function 
is obtained by forcing  
$\tilde{\epsilon}_{i,\uparrow}=\tilde{\epsilon}_{i,\downarrow}$, while, to 
discuss magnetism, we allow the wave function to break spin-rotational symmetry
with $\tilde{\epsilon}_{i,\uparrow} \ne \tilde{\epsilon}_{i,\downarrow}$. 
${\cal P}_G= \exp{\left[ \sum_i\, g_i\, n_i^2 \right]}$ is a Gutzwiller
correlator that depends upon the {\it site-dependent} parameters $g_i$'s, while
${\cal J}= \exp{\left[ 1/2 \sum_{i\not = j}\, v_{i,j} (n_i-1)(n_j-1) \right]}$ 
is a Jastrow factor. The latter one spatially correlates valence fluctuations, 
$\delta n_i=\langle n_i-1\rangle \not = 0$, on different sites, binding those 
with $\delta n_i\,\delta n_j<0$ and unbinding those with 
$\delta n_i\,\delta n_j>0$. This fact has been shown to be crucial to describe 
a Mott transition in clean systems.~\cite{capello,capello2}
We shall assume that $v_{ij}$ is translationally invariant, which makes the 
numerical calculations feasible but neglects any clustering effects.
All the parameters contained in the variational wave function $|\Psi \rangle$, 
i.e., $\tilde{\epsilon}_{i,\sigma}$, $g_i$, and $v_{i,j}$,~\cite{notet} are 
optimized to minimize the variational energy by using the Monte Carlo 
technique of Ref.~\cite{sorella}.

\begin{figure}
\includegraphics[width=0.50\textwidth]{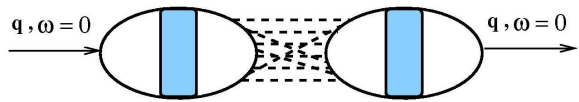}
\caption{\label{fig:diagram} 
(Color online) Density-density fluctuations 
$N_q^{\rm disc} = \overline{\langle n_q \rangle \langle n_{-q} \rangle}$. 
Dotted lines denote impurity averages, and the squares indicate vertex 
corrections that include both interaction and impurity insertions. Continuous 
lines are fully corrected Green's functions.}
\end{figure}

As discussed in Refs.~\cite{capello,capello2} for clean systems, it is possible
to discriminate variationally metals from Mott insulators by looking to the 
equal-time density-density structure factor  
$N_q=\langle \Psi| n_q \,n_{-q}|\Psi \rangle/ \langle \Psi|\Psi \rangle$,
where $n_q$ is the Fourier transform of the electron density $n_i$.
Indeed, $N_q \sim |q|$ implies the existence of gapless modes, while 
$N_q \sim |q|^2$ indicates that charge excitations are gapped. 
Moreover, there is a tight connection between the long-wave-length behavior 
of $N_q$ and the Fourier transform of the Jastrow factor $v_q$, namely
$v_q \sim 1/|q|$ for a metal and $v_q \sim 1/|q|^2$ for an 
insulator.~\cite{capello,capello2} This distinction should equally work 
in~(\ref{Ham}) after disorder average. However, particular care must be taken 
to interpret $N_q$ in a disordered system, where the structure factor includes 
a {\it disconnected} term, 
$N_q^{\rm disc} = \overline{\langle n_q \rangle \langle n_{-q} \rangle}$ 
(where the quantum average is taken at fixed disorder configuration and 
the overbar indicates the disorder average) as well as a connected one, i.e.,
$N_q^{\rm conn} = N_q - N_q^{\rm disc}$. For a clean system, the disconnected 
term gives rise to the elastic scattering peaks at $q$ equal to the reciprocal 
lattice vectors, the Bragg reflections. On the contrary, in the presence of 
disorder $N_q^{\rm disc}$ is finite for any finite momentum $q$.~\cite{gold} 
The diagrammatic representation of $N_q^{\rm disc}$ is shown in 
Fig.~\ref{fig:diagram} and one can realize that, for $q\to 0$, it reduces to 
the electron compressibility fluctuations. For non-interacting electrons, 
$N_q^{\rm disc}$ is finite for $q \to 0$, whereas $N_q^{\rm conn} \sim |q|$, 
indicating the absence of a gap in the spectrum of charge-density 
fluctuations.~\cite{gold}

\begin{figure}
\includegraphics[width=0.50\textwidth]{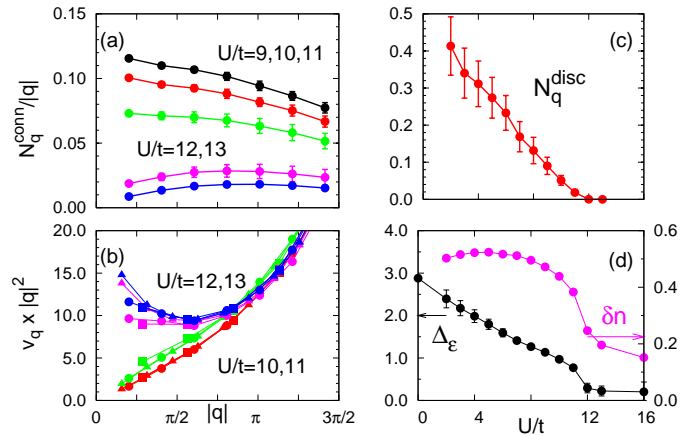}
\caption{\label{fig:param} 
(Color online) (a) Connected term of the density-density correlation 
function $N_q^{\rm conn}$ divided by $|q|$. (b) Jastrow parameters $v_q$ 
multiplied by $|q|^2$. (c) Disconnected term of the density-density correlation 
function $N_q^{\rm disc}$ as a function of $U$. (d) Fluctuations of the on-site
variational energies $\Delta_\epsilon$ and of the local densities.
All calculations have been done for $D/t=5$.} 
\end{figure}

We start our analysis with the case of nearest-neighbor hopping only by using 
a paramagnetic wave function, namely imposing
$\tilde{\epsilon}_{i,\uparrow}=\tilde{\epsilon}_{i,\downarrow}$.
In Fig.~\ref{fig:param}, we show the variational $N_q^{\rm conn}$ and the 
Fourier transform of the optimized Jastrow potential $v_q$ for different 
values of the interaction $U$ and $D/t=5$ (we take such a large value of $D$ 
in order to have a localization length that, at $U=0$, is smaller than 
the numerically accessible system sizes). A clear change in the behavior of 
these quantities is observed at $U_c^{\rm Mott}/t=11.5 \pm 0.5$. For small 
values of the electron interaction, $N_q^{\rm conn} \sim |q|$ and 
$v_q \sim 1/|q|$, whereas $N_q^{\rm conn} \sim |q|^2 $ and $v_q \sim 1/|q|^2$ 
in the strong-coupling regime. The latter behavior is symptomatic of 
the presence of a charge gap hence of a Mott insulating 
behavior.~\cite{capello} We notice that, for the clean case $D=0$ and within 
the same approach, a metal-insulator transition at $U_c^{\rm Mott}=8.5 \pm 0.5$
was found,~\cite{capello-prb} indicating that disorder competes with $U$ and 
pushes the Mott transition to higher values of $U/t$. 
It should be emphasized that, with respect to the clean system, for 
$U<U_c^{\rm Mott}$, $N_q^{\rm conn} \sim |q|$ is not associated to a metallic 
behavior but only to a gapless spectrum, also characteristic of an Anderson 
insulator. Remarkably, we find that the Mott and Anderson insulators can also 
be  discriminated through the behavior of the  $\lim_{q \to 0} N_q^{\rm disc}$.
In Fig.~\ref{fig:param} we plot this quantity for different values of $U$, 
demonstrating that it is finite in the Anderson insulator, whereas it vanishes 
in the Mott phase. This identifies a simple and variationally accessible order 
parameter for the Anderson-Mott transition.

Even though within this approach we cannot access dynamical 
quantities like DC conductivity, hence we can not address the question of a 
possible stabilization of a conducting phase with moderate Coulomb 
repulsion,~\cite{scalettar} we note that the linear slope of $N_q^{\rm conn}$ 
has a non-monotonic behavior as a function of $U$, showing a peak for 
$U/t \sim 7$ that indicates an accumulation of low energy states around the 
Fermi energy. The same qualitative behavior is also present in the 
fluctuations of the local densities, 
$\delta n^2=1/N \sum_i (\langle n_i^2 \rangle - \langle n_i \rangle^2)$.
Though the single-particle eigenstates of the variational 
Hamiltonian may have a very long localization length, because of the 
suppression of the effective on site disorder $\tilde{\epsilon}_i$, yet this 
length is still finite in two dimensions hence the many-body wave function 
$|\Psi\rangle$ always describes an Anderson insulator below the Mott 
transition. Indeed, as shown in Fig.~\ref{fig:param}, the fluctuations of the 
on-site variational disorder $\Delta_\epsilon^2=
1/N \sum_i \tilde{\epsilon}_i^2 - (1/N \sum_i \tilde{\epsilon_i})^2$ 
are always finite, though sizably renormalized by the electron interaction $U$. 

\begin{figure}
\includegraphics[width=0.50\textwidth]{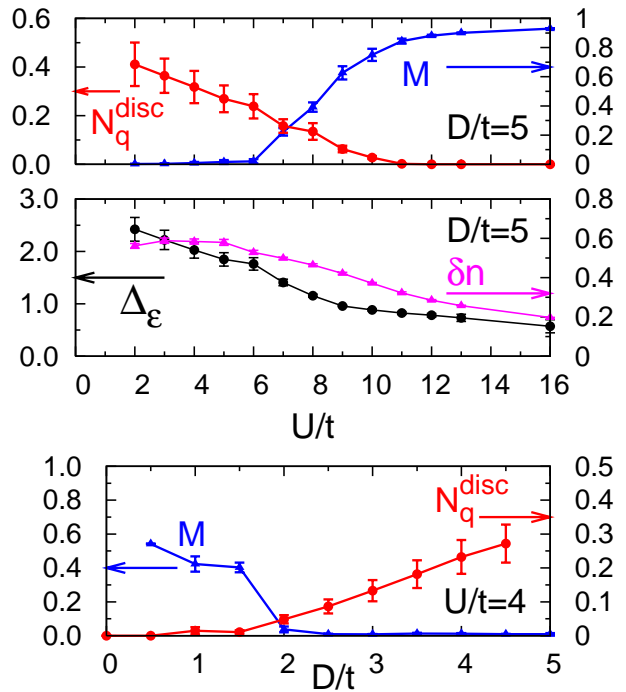}
\caption{\label{fig:magnetic} 
(Color online) Staggered magnetization $M$ for $Q=(\pi,\pi)$ and 
compressibility fluctuations $N_q^{\rm disc}$ as a function of $U$ for 
disorder $D/t=5$ (upper panel) and as a function of $D$ for $U/t=4$ 
(bottom panel). Fluctuations of the on-site variational energies 
$\Delta_\epsilon$ and of the local densities (middle panel). 
Calculations have been done for $N=98$ and error-bars indicate the average 
over different realizations of disorder.}
\end{figure}

\begin{figure}
\includegraphics[width=0.50\textwidth]{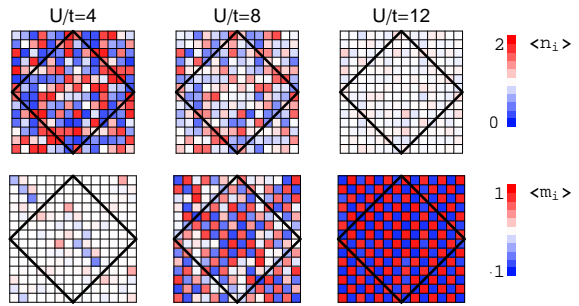}
\caption{\label{fig:tappeti} 
(Color online) Local density $\langle n_i \rangle$ (upper panels) and  
local magnetization $\langle m_i \rangle$ (lower panels) for a given disorder 
realization with $D/t=5$ and different values of $U/t$. The black contour 
shows the elementary cell of the lattice which it is repeated to mimic the 
infinite lattice with periodic boundary conditions.}
\end{figure}

Let us now move to the more interesting case in which we allow magnetism in 
the variational wave function, which amounts to permit  
$\tilde{\epsilon}_{i,\uparrow} \ne \tilde{\epsilon}_{i,\downarrow}$.
In this case the ground state may acquire a finite {\it local} magnetization 
on each site $m_i=n_{i,\uparrow}- n_{i,\downarrow}$. A magnetically ordered 
phase will have a finite value of the total magnetization 
$M=1/N \sum_j e^{\imath R_j Q} m_j$ for a suitable momentum $Q$, like for 
instance $Q=(\pi,\pi)$ for the N\'eel state.
In the presence of disorder, a finite value $U_c^{\rm AF}$ is needed to have 
long-range antiferromagnetic order. We find that, also in presence of a
small $t^\prime$, $U_c^{\rm AF}<U_c^{\rm Mott}$, giving rise to an extended 
region with antiferromagnetic order and finite compressibility (i.e., a 
vanishing charge gap). These results are in agreement with previous mean-field 
calculations.~\cite{heidarian,janis,janis-bis} In Fig.~\ref{fig:magnetic}, we 
show the results for $t^\prime=0$ either by fixing $D/t=5$ and varying $U$ 
(for which $U_c^{\rm AF}/t=6.5 \pm 0.5$ and $U_c^{\rm Mott}/t=10.5 \pm 0.5$)
or by fixing $U/t=4$ and changing $D$ (for which $D_c^{\rm Mott}/t=1 \pm 0.5$ 
and $D_c^{\rm AF}/t=2.5 \pm 0.5$). We note that the onset of antiferromagnetism 
is preceded by a magnetically disordered phase (i.e., $M=0$) in which 
local moments appear. In Fig.~\ref{fig:tappeti}, the pattern of the local 
density $\langle n_i \rangle$ and local magnetization $\langle m_i \rangle$ 
are shown for a typical realization of disorder. For $U/t=4$, the ground state 
is an Anderson insulator with a large number of empty and doubly occupied 
sites with $m_i \sim 0$. However, some sites have finite magnetization, but 
they are not spatially correlated hence long-range magnetism is absent. 
We interpret these magnetic sites as local moments. When the electron 
interaction $U$ increases, the number of magnetic sites increases rapidly and 
the local moments eventually display the typical staggered pattern of N\'eel 
order. Nevertheless, charge excitations are still gapless, with
$N_q^{\rm conn} \sim |q|$. For $U/t=12$ the system is a gapped insulator with 
antiferromagnetic order and a vanishing compressibility. Variationally, the 
charge gap opens by the combined effect of the Jastrow correlations, i.e., 
$v_q \sim 1/|q|^2$, and the finite antiferromagnetic gap in the mean-field 
Hamiltonian (due to staggered $\tilde{\epsilon}_{i,\sigma}$'s).

\begin{figure}
\includegraphics[width=0.50\textwidth]{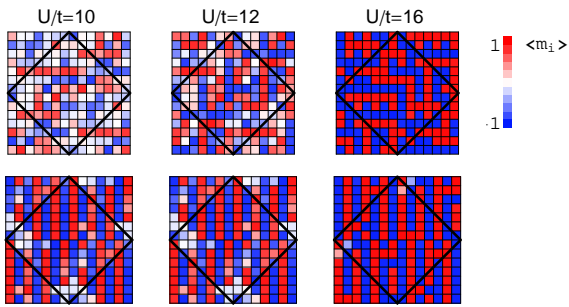}
\caption{\label{fig:tappetitp} 
(Color online) Local magnetization $\langle m_i \rangle$ for the best 
variational state (lower panels) and for a metastable solution (upper panels)
for a given disorder configuration with $D/t=5$ and $t^\prime/t=1$.}
\end{figure}

In the presence of a large frustrating hopping $t^\prime/t \gtrsim 0.9$ we 
find evidences of a spin glass behavior. In the large $U$ regime, the optimal 
wave function displays magnetic long-range order with $Q=(\pi,0)$ or $(0,\pi)$. 
However, the energy landscape contains other local minima very close in energy 
in which most of the sites of the lattice have a net magnetization but an 
overall vanishing magnetic order, a ``glassy'' spin patterns, 
see Fig.~\ref{fig:tappetitp}. These solutions are incompressible, i.e., 
$N_q^{\rm disc} \sim 0$ and, therefore, may be viewed as disordered Mott 
insulators. By decreasing the interaction strength $U$, these metastable states
turn compressible, still having a large number of local moments. However, 
the actual variational minimum shows, as before, a Mott transition from a Mott 
to an Anderson insulator, both magnetically ordered, followed, at lower $U$, 
by a further transition into a paramagnetic Anderson insulator. The only role 
of $t^\prime$ is to shrink the region in which a magnetic Anderson insulator 
is stable.

In conclusion, we have shown that a relatively simple variational wave function
is able to describe the Anderson-insulator to Mott-insulator transition in two 
dimensions. In the paramagnetic sector, this phase transition is continuous, 
in agreement with dynamical mean field theory.~\cite{DMFT,hofstetter}
When spontaneous spin symmetry breaking is allowed, we find two successive 
phase transition, the first from a paramagnetic Anderson insulator to a 
magnetic one, followed by a transition from a magnetic Anderson insulator to a 
magnetic Mott insulator. Upon increasing frustration, the stability region of 
the magnetic Anderson insulator decreases. In general, the paramagnetic 
Anderson insulator develops local magnetic moments, but we do not find 
any evidence of a truly metallic behavior induced by interaction.    

We acknowledge partial support from CNR-INFM.


\begin{thebibliography}{99}
\bibitem{notet} In principle, one could also allow for site-dependent hopping
   amplitudes in the mean-field Hamiltonian ${\cal H}_{MF}$, however we
   checked that this further variational freedom does not qualitatively modify
   the final results.
\end{thebibliography}
\end{document}